\documentclass[prl,noshowkeys,showpacs,twocolumn,superscriptaddress]{revtex4}
\usepackage{graphicx}
\usepackage{amsmath}

\def\gap{\;\rlap{\lower 2.5pt
 \hbox{$\sim$}}\raise 1.5pt\hbox{$>$}\;}
\def\lap{\;\rlap{\lower 2.5pt
   \hbox{$\sim$}}\raise 1.5pt\hbox{$<$}\;}
\def\gsim{\;\rlap{\lower 2.5pt
 \hbox{$\sim$}}\raise 1.5pt\hbox{$>$}\;}
\def\lsim{\;\rlap{\lower 2.5pt
   \hbox{$\sim$}}\raise 1.5pt\hbox{$<$}\;}
\def\msun{{\rm\,M_\odot}}

\def\spose#1{\hbox to 0pt{#1\hss}}
\def\lta{\mathrel{\spose{\lower 3pt\hbox{$\mathchar''218$}}
     \raise 2.0pt\hbox{$\mathchar''13C$}}}
\def\gta{\mathrel{\spose{\lower 3pt\hbox{$\mathchar''218$}}
     \raise 2.0pt\hbox{$\mathchar''13E$}}}
\newcommand{\be}{\begin{equation}}
\newcommand{\ee}{\end{equation}}

\newcommand{\ls}{\mathrel{\raise1.16pt\hbox{$<$}\kern-7.0pt 
\lower3.06pt\hbox{{$\scriptstyle \sim$}}}}         
\newcommand{\gs}{\mathrel{\raise1.16pt\hbox{$>$}\kern-7.0pt 
\lower3.06pt\hbox{{$\scriptstyle \sim$}}}}         

\long\def\comment#1{}

\def\mh{M_{\bullet}}
\def\msun{M_{\odot}}
\def\fun#1#2{\lower3.6pt\vbox{\baselineskip0pt\lineskip.9pt
  \ialign{$\mathsurround=0pt#1\hfil##\hfil$\crcr#2\crcr\sim\crcr}}}
\def\lap{\mathrel{\mathpalette\fun <}}
\def\gap{\mathrel{\mathpalette\fun >}}
\newcommand{\ba}{\begin{eqnarray}}
\newcommand{\ea}{\end{eqnarray}}

\begin{document}
\bibliographystyle{apsrev.bst}
\title{Time-Dependent Models for Dark Matter at the Galactic Center}
\author{Gianfranco Bertone}
\affiliation{NASA/Fermilab Theoretical Astrophysics Group,
Batavia IL 60510, USA}
\author{David Merritt}
\affiliation{Department of Physics, Rochester Institute of Technology,
Rochester, NY 14623, USA}

\begin{abstract}
The prospects for indirect detection of dark matter at the Galactic center 
with gamma-ray experiments like the space telescope GLAST, and 
Air Cherenkov Telescopes like HESS, CANGAROO, MAGIC and VERITAS,
depend sensitively on the mass profile within the inner parsec. 
We calculate the distribution of dark matter on sub-parsec scales by 
integrating the time-dependent Fokker-Planck equation, including the 
effects of self-annihilations, 
scattering of dark matter particles by stars,
and capture in the supermassive black hole. 
We consider a variety of initial dark matter distributions, 
including models with very high densities (``spikes'') near the black hole,
and models with ``adiabatic compression'' of the baryons. 
The annihilation signal after $10^{10}$ yr is found to be substantially 
reduced from its initial value, but in dark matter models with an initial 
spike, order-of-magnitude enhancements can persist compared with the rate in 
spike-free models.

\end{abstract}
\pacs{
\hspace{4.8cm} FERMILAB-PUB-05-013-A}
\maketitle

There is compelling evidence that the matter density of the Universe 
is dominated by some sort of non-baryonic, ``dark'', matter, 
the best candidates being weakly interacting massive particles
\cite{Bertone:04,Bergstrom:2000pn}. 
Numerical $N$-body simulations suggest dark 
matter (DM) density profiles following broken power laws, 
$\rho\propto r^{-\gamma}$,
with $\gamma\approx 3$ in the outer parts of halos and 
$1\lap\gamma\lap 1.5$ 
(``cusps'') inside the Solar circle.
Although these profiles reproduce with sufficently good accuracy the 
observed properties of galactic halos on large scales, as inferred by 
rotation curves, little is known about the DM distribution
on smaller scales, where the gravitational potential is dominated by 
baryons. 
The situation at the Galactic center (GC) is further complicated by the 
presence of a supermassive black hole (SBH), with mass $\sim 10^{6.5}$  
M$_\odot$~\cite{Schoedel:03}, whose sphere of gravitational influence 
extends out to $\sim 1$ pc. 

The prospects for indirect detection depend crucially on the
distribution of DM within this small region. The flux of gamma-rays 
from the GC, from the annihilation of DM particles of mass $m$ and
annihilation cross section in the non-relativistic limit $\sigma v$, can
be written:
\be
\Phi_{i}(\Delta\Omega, E) \simeq \Phi_0 \frac{dN_i}{dE} 
\left( \frac{\sigma v}{\langle\sigma v\rangle_{th}} \right)
\left(\frac{1\rm{TeV}} 
{m} \right)^2 \overline{J}_{\Delta\Omega} \Delta\Omega
\label{final}
\ee
where $\Phi_0=5.6\times10^{-12}\,\rm{cm}^{-2} \rm{s}^{-1}$ and 
and $\langle\sigma v\rangle_{th}=3\times 10^{-26}{\rm cm}^3 {\rm s}^{-1}$
is the value of the thermally averaged cross section at decoupling that 
reproduces the observed cosmological abundance of Dark Matter (although 
in presence of resonance effects like co-annihilations, the correct 
relic abundance can be achieved with smaller cross sections). For more
details and a review on DM candidates and detection see e.g. 
Refs.\cite{Bertone:04,Bergstrom:2000pn}.  
$\overline{J}_{\Delta\Omega}$ is a factor containing all the information 
on the DM profile~\cite{Bergstrom:1997fj}:
\begin{equation}
{\overline J}_{\Delta\Omega}\equiv K (\Delta\Omega)^{-1}\int_{\Delta\Omega}d\psi \int_{\psi} \rho^2 dl,
\end{equation}
where $dl$ is the distance element along the line of sight at angle
$\psi$ with respect to the GC, $\Delta\Omega$ is the 
solid angle of the detector, and $K$ is a normalizing factor,
$K^{-1}=(8.5 {\rm kpc})(0.3 {\rm GeV}/{\rm cm}^3)^2$.
We denote by $\overline{J}_5$ and $\overline{J}_3$ the values of $\overline{J}$
when $\Delta\Omega=10^{-5}$sr and $10^{-3}$sr respectively;
the former is the approximate field of view of the 
detectors in GLAST \cite{GLAST} and in atmospheric Cerenkov telescopes 
like VERITAS \cite{VERITAS} and HESS \cite{HESS}, 
while the larger angle corresponds approximately
to EGRET \cite{EGRET}. 
DM densities that rise more steeply than $\rho\propto r^{-3/2}$
near the GC imply formally divergent values of $\overline{J}$,
hence the predicted flux of annihilation products can depend sensitively
on any physical processes that modify the DM density on subparsec scales.
Although the analysis of DM indirect detection is usually performed
under simplifying assumptions on the DM profile -- extrapolating
the results of numerical simulations with power-laws down to subparsec 
scales -- several dynamical processes may influence 
the distribution of DM at the GC, including the gravitational
force from the SBH \cite{Gondolo:99}, 
adiabatic compression of baryons \cite{Blumenthal:86}, 
and heating of the DM by stars \cite{Merritt:04}. 

Here, we focus on the evolution of the annihilation signal
due to two physical processes that are almost certain to
strongly influence  the form of the DM density profile near the GC:
DM self-annihilations; and gravitational interactions between
DM particles and stars.
Both processes act on a similar time scale ($\sim 10^9$ yr) to
modify $\rho(r)$ on the sub-parsec scales that are most
relevant to the indirect detection problem.
While these two mechanisms both tend to lower the DM density, 
we find that interestingly high densities can persist 
over a particular range of $(m,\sigma v)$ values.
The time-dependent profiles discussed here may also have 
important consequences for the prospects of observing an extra-galactic
gamma-ray background.

Let $f({\bf r},{\bf v}, t)$ be the mass density of DM
particles in phase space and $\rho(r,t)$ their
configuration-space density, with $r$ the distance from the
GC, i.e. the distance from the SBH.
We assume an isotropic velocity distribution,
$f({\bf r}, {\bf v},t)=f(E,t)$, where $E\equiv -v^2/2 +\phi(r)$
is the binding energy per unit mass and $-\phi(r)$ is the
gravitational potential, which includes contributions
from the stars in the Galactic bulge and from the SBH.
We assume that $\phi$ is fixed in time, i.e. that the mass of 
the SBH has not changed since the epoch of cusp formation,
and that the stellar distribution has also not evolved.
The first assumption is commonly made based on the observed,
very early formation of massive black holes (e.g. \cite{Fan:04}).
The second assumption is motivated by the expectation 
that the stars should reach a collisional steady state
around the SBH,
the so-called ``Bahcall-Wolf'' solution \cite{BW:76},
in a time of $\sim 10^9$ yr.
The observed stellar distribution at the Galactic
center, $\rho_\star \sim r^{-1.4}$ \cite{Genzel:03},
is slightly shallower than the Bahcall-Wolf solution
for a single population but is generally believed to 
be consistent with a collisional steady state given 
uncertainties about the stellar mass spectrum
\cite{Alexander:99}.
We accordingly set
$\rho_\star(r) \propto r^{-1.4}$ and
fix its normalization to match the observed stellar
density at $\sim 1$ pc from the SBH \cite{Genzel:03}.
The stellar phase-space mass density $f_\star(E)$ is then
uniquely determined by $\rho_\star(r)$ and $\phi(r)$ via
Eddington's formula \cite{Eddington:16,GS:99}.

We describe the evolution of $f$
via the orbit-averaged Fokker-Planck equation including
loss terms (e.g. \cite{Spitzer:87}):
\begin{subequations}
\begin{eqnarray}
&&{\partial f\over\partial t} = 
{1\over 4\pi^2p}{\partial \over\partial E}
\left[D_{EE}{\partial f\over\partial E}\right] 
-f(E,t)\nu_{\rm coll}(E) \nonumber \\
&&\ \ \ \ \ \ \ \ - f(E,t)\nu_{\rm lc}(E), \label{eq:fpa}\\
&&D_{EE}(E) = 64\pi^4G^2{\overline m}_\star\ln\Lambda\times  \nonumber \\ 
&&\left[q(E)\int_{-\infty}^E dE'f_\star(E') + 
\int_E^\infty dE' q(E')f_\star(E')\right].
\label{eq:fpb}
\end{eqnarray}
\end{subequations}
Here $p(E) = 4\sqrt{2}\int_0^{r_{max}(E)} dr r^2 \sqrt{\phi(r)-E}$
is the phase space volume accessible per unit of energy,
$p(E) = -\partial q/\partial E$, and
$\ln\Lambda$ is the Coulomb logarithm \cite{Spitzer:87}.
We have assumed that the spectrum of stellar masses
$n(m_\star)dm_\star$ is independent of radius;
then ${\overline m}_\star= \langle m_\star^2\rangle/\langle m_\star\rangle$
\cite{Merritt:04}.
 
The first term on the right-hand side of Eq. 3a
describes the diffusion of DM particles in energy
space due to heating via gravitational encounters with stars.
Near the SBH, the characteristic heating time is
nearly independent of energy and radius and is given
approximately by 
$T_{\rm heat} = 1.25\times 10^9 {\rm yr}      
\tilde{\mh}^{-1/2} \tilde{r_h}^{3/2} \tilde{m_\star}^{-1}$
with $\tilde{\mh}=\mh/(3\times 10^6\msun)$;
$\tilde{r_h}=r_h/(2{\rm pc})$, where $r_h$ is 
the ``gravitational influence
radius'' of the SBH, defined as the radius 
of the sphere containing a mass in stars equal to
twice $\mh$; and $\tilde{m_\star}=\overline{m}_\star/\msun$
\cite{Merritt:04}.
In what follows we set $\tilde{\mh}=\tilde{r_h}=\tilde{m_\star}=1$ 
and define $\tau\equiv t/T_{\rm heat}$;
the age of the Galactic bulge, $\sim 10$ Gyr,
then corresponds to $\tau \approx 10$.
(The most recent estimates of $\mh$ are slightly higher
\cite{Ghez:05}; we adopt $\tilde\mh=1$ for consistency
with earlier work \cite{Merritt:04}.)

The collision term $\nu_{\rm coll}$ has two
potential contributors: self-annihilations, and
interaction of DM particles with
baryons.
The self-annihilation term is given locally by 
$\nu=m^{-1}\rho \sigma v$.
The orbit-averaged rate $\nu_{\rm coll}$ that appears in 
Eq. 3a is 
\begin{equation}
\nu_{\rm coll}(E) = {\int \nu r^2 v(r,E) dr\over \int r^2 v(r,E) dr}
\end{equation}
where $v(r,E)=\sqrt{2\left(\phi(r)-E\right)}$ and the integrals
are from $0$ to $r_{\rm max}(E)$.
Expressing $\rho$ in terms of $f$, we can write the orbit-averaged
self-annihilation term as
\begin{subequations}
\begin{eqnarray}
& &p(E)\nu_{\rm coll}(E) = 32\pi (\sigma v) m^{-1}\times \nonumber 
\\
& &\left[\int_0^E dE'f(E')C(E,E') \int_E^\infty dE'f(E')C(E',E)\right],
\\
& &C(E,E') \equiv \int_0^{\phi^{-1}(E)} dr r^2\sqrt{\phi(r)-E}\sqrt{\phi(r)-E'}.
\end{eqnarray}
\end{subequations}
Self-annihilations limit the DM density roughly to 
$\rho\approx m/ (\sigma v) t$ \cite{Berezinsky:92,Gondolo:99};
for $m=50$ GeV, $\sigma v=10^{-26}$ cm$^3$s$^{-1}$
and $t=10$ Gyr, $\rho\lap 2\times 10^6\msun{\rm pc}^{-3}$,
which would imply that self-annihilations are important at $r\lap 10^{-3}$ pc
if $\rho\sim r^{-3/2}$ and $\rho(r_h) = 100\msun$ pc$^{-3}$.


The final loss term in Eq. 3 represents
scattering of DM particles into the SBH \cite{Merritt:04}.
This term, which we include, is important at radii $r\lap r_h$.
The loss rate varies only logarithmically with the
SBH's capture radius, which we set to $2G\mh/c^2$.

In what follows, we assume $\tilde m_\star = 1 \msun$,
consistent with our limited knowledge of the stellar mass
spectrum near the Galactic center \cite{Merritt:04}.
We note that both the first and third terms on the right hand
side of eq. 3a depend in the same way on $\tilde m_\star$; thus, 
varying $\tilde m_\star$ has the effect of changing the relative time
scales for DM-star scattering and self-annihilations.
Since changing $\sigma v$ has the same effect,
we do not vary $\tilde m_\star$ in what follows.

Eq. 3 was advanced in time via a backward
differentiation scheme coupled with the method of lines
to reduce the partial differential equation to a system
of ODEs \cite{Berzins:89}
A variable time step was employed, such that the fractional
change in $f$ in one time step was less than $1\%$ at
every value of $E$.
For very high values of $\sigma v/m$,
the initial $f$ was truncated such that
the annihilation time was never shorter than $\sim 10^6$ yr.

We adopted a wide range of initial conditions for the
DM distribution (Table I).
Baryon-free simulations of DM clustering suggest
a power-law distribution in the inner parts of galaxies,
$\rho\propto r^{-\gamma_{c}}$, a ``cusp,'' with $\gamma_c\approx 1$ 
\cite{NFW:96} (these models are labelled ``N'' in Table I).
The most recent simulations \cite{Navarro:03,Reed:05} 
(see also Refs.~\cite{Fukushige:2003xc,Diemand:2004wh})
suggest a power-law index that varies slowly with radius,
but the normalization and slope of these models at
$r\approx r_h$ are essentially identical to those of 
models with an unbroken, $\rho\propto r^{-1}$ power law
inward of the Sun.
We took $R_\odot=8.0$ kpc for the radius of the Solar
circle \cite{Eisenhauer:03}.

Since the total mass budget of the inner Galaxy
is dominated by baryons, the DM distribution
is likely to have been influenced by the 
baryonic potential and its changes over time.
The ``adiabatic-growth'' model \cite{Blumenthal:86}
posits that the baryons contracted quasi-statically
and symmetrically within the pre-existing
DM halo, pulling in the DM and increasing its density.
When applied to a DM halo with initial $\gamma_c\approx 1$,
the result is a halo profile with $\gamma_c\approx 1.5$ 
inward of $R_\odot$ and an increased density at $R_\odot$ 
\cite{Edsjo:2004pf,Prada:2004pi,Gnedin:2004}.
Adiabatically contracted halo models are labelled ``A'' in
Table I.
Alternatively, strong departures from spherical symmetry
during galaxy formation might have resulted in a
{\it lower} central DM density.
For instance, the DM density following a merger is a weak power law, 
$\rho\sim r^{-\gamma_{in}}$, $\gamma_{in}\approx 0.5$, 
inside a radius $r_c\approx 10-100 r_h$ \cite{Merritt:02,Ullio:2001fb}.
Models with the subscript ``c'' in Table I
have $\rho\propto r^{-1/2}$ inside a radius $r_c=10$pc.

We also considered modified versions of each of these
DM profiles that included a density ``spike'' around the SBH;
these models are denoted by the subscript ``sp'' in Table I.
The inner DM density in the spike models follows 
the steeper power law that would
result from gradual growth of the SBH to its current mass
at a fixed location.
We set $\rho=\rho(r_b)(r/r_b)^{-\gamma_{sp}}$ for $r\lap r_b$
with $\gamma_{sp} = 2 + 1/(4-\gamma)$ and $\gamma$
the power-law index of the core or cusp,
and $r_b = 0.2 r_h$ \cite{Gondolo:99,Merritt:04b}.
It is unclear whether spikes can survive at the
centers of all galactic halos, since dynamical effects 
such as off-center formation of the SBH and binary black 
hole mergers would tend to destroy high density regions
~\cite{Ullio:2001fb,Merritt:02}. However, the Milky Way 
is unlikely to have experienced a ``major merger''
(a merger with another galaxy of comparable mass) 
in the last 10 Gyr,
and the existence of a stellar ``cusp'' \cite{Genzel:03} further 
strengthens the case for a dark matter spike at the Galactic 
center \cite{Silk:02}. 

\begin{table}
\caption{\label{Table1}Properties of the halo models. ``N''
and ``A'' stand for NFW and adiabatically contracted profiles,
respectively.
The subscripts ``c'', ``sp'' are for profiles with core
and spike respectively. Core radius $r_c$ is in units of $r_h$.
Density at $R_\odot$ is in units of GeV cm$^{-3}$.
$\overline{J}_3$ and $\overline{J}_5$ are
values of $J$ averaged over windows of solid angle
$10^{-3}$ sr and $10^{-5}$ sr respectively and normalized
as described in the text.
The final two columns give $\overline{J}$ in evolved models
for $\sigma v=0$ (no annihilations),
and for $\left(\sigma v, m\right)=
\left(3\times 10^{-26}\mathrm{cm}^3\mathrm{s}^{-1}, 50 \mathrm{GeV}\right)$
(maximal annihilation rate), respectively.}
\begin{tabular}{|c|cccc|ccc|}
\hline
\hline
 & & & & & & $\log_{10} \overline{J}_3$ ($\overline{J}_5$) &  \\
 & $\gamma_c$ & $\gamma_{sp}$ & $r_c$ & $\rho(R_\odot)$ & 
$\tau=0$ & $\tau=10$ & $\tau=10$ \\
\hline
$N$ & 1.0 & -- & -- & 0.3 & 2.56~(3.51) & 2.56~(3.50) & 2.56~(3.50)\\
$N_c$ & 1.0 & -- & 10 & 0.3 & 2.54~(3.33) & 2.54~(3.33) & 2.54~(3.33)\\
$N_{sp}$ & 1.0 & 2.33 & -- & 0.3 & 9.21~(11.2) & 3.86~(5.84) & 2.56~(3.52)\\
$N_{c,sp}$ & 1.0 & 2.29 & 10 & 0.3 & 6.98~(8.98) & 2.61~(3.88) & 2.54~(3.33)\\
\hline
$A$ & 1.5 & -- & -- & 0.5 & 5.80~(7.75) & 5.26~(7.03) & 5.23~(6.98)\\
$A_c$ & 1.5 & -- & 10 & 0.5 & 4.96~(6.27) & 4.96~(6.27) & 4.96~(6.27)\\
$A_{sp}$ & 1.5 & 2.40 & -- & 0.5 & 14.8~(16.8) & 9.25~(11.3) & 5.25~(7.02)\\
$A_{c,sp}$ & 1.5 & 2.29 & 10 & 0.5 & 9.99~(12.0) & 5.21~(6.96) & 4.96~(6.27)\\
\hline
\hline
\end{tabular}
\end{table}

In order to evaluate the influence of annihilations on the
evolution of the DM profile, we first investigated two
extreme cases in the framework of typical DM
candidates like neutralinos or Kaluza-Klein particles~\cite{Bertone:04}.
In the first extreme case, in order to maximize the ratio 
$\sigma v/m$,
we assumed a cross section 
$\sigma v=\langle\sigma v\rangle_{\rm{th}} = 3\times 10^{-26}$ cm$^3$ s$^{-1}$
and a mass of 50 GeV. Higher values
of the annihilation cross section, though possible, would imply a low
relic density, making the candidate a subdominant component of the
DM in the Universe, a case we are not interested in here. 
The lower limit on the mass strictly applies only to neutralinos in 
theories with gaugino and sfermion mass unification at the GUT 
scale~\cite{Eidelman:2004wy}, while the limit on the mass of KK particles 
is higher. 
The second extreme case assumes no annihilations, as in the
limit of very small cross sections, or very heavy particles.
Table I gives values of $\overline{J}_3$ and $\overline{J}_5$ at $\tau=10$ for each of
our DM models and for both of the extreme particle physics
models.
The $\overline{J}$-values depend appreciably on the particle physics model
only when the initial DM density has a ``spike''
around the SBH; in other cases the central density is too low
for annihilations to affect $\overline{J}$.
Particularly in the case of maximal $\sigma v$,
the final $\overline{J}$ values are found to be modest, 
$\log_{10}\overline{J}_3\lap 5.3$ and $\log_{10}\overline{J}_5\lap 7.0$,
compared with the much larger values at $\tau=0$ in the presence
of spikes.

\begin{figure}
\includegraphics[height=12cm]{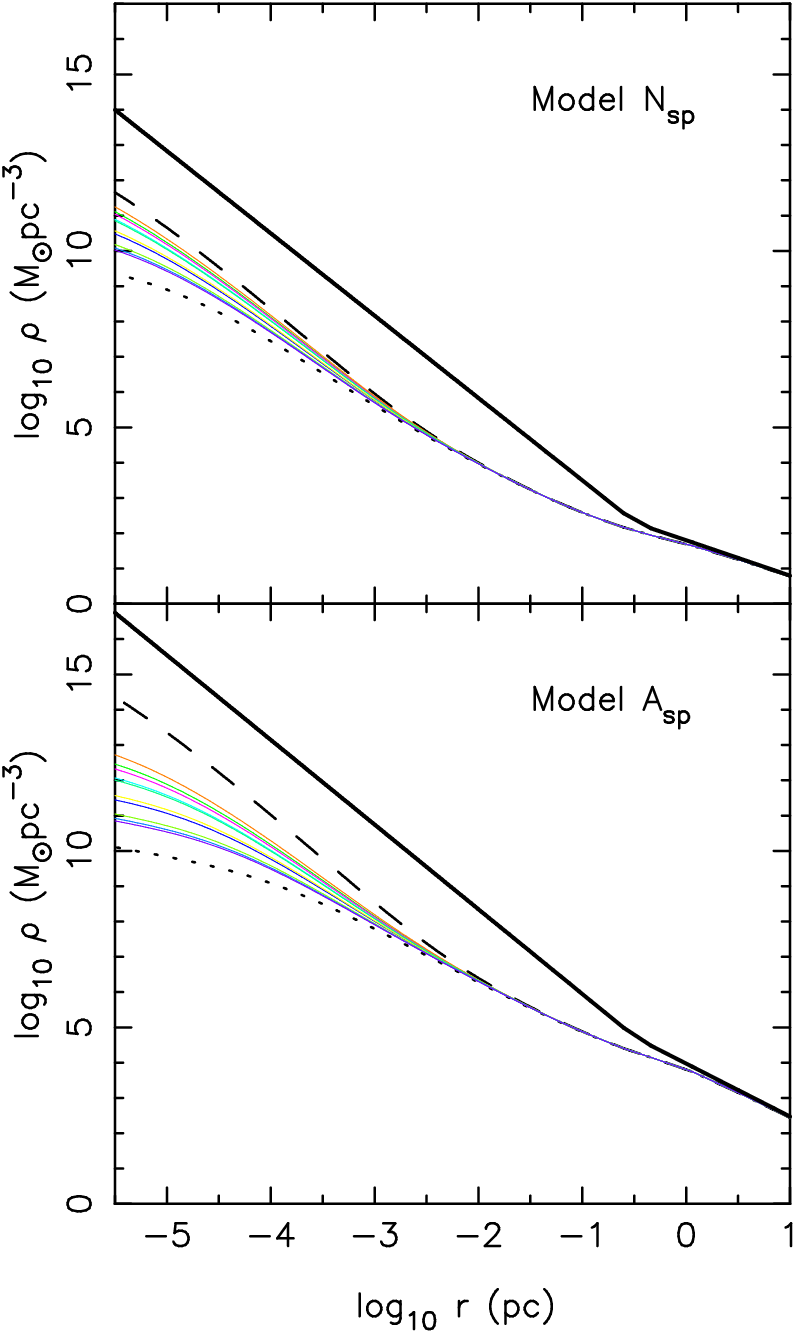}
\caption{Evolved DM density profiles at $\tau=10$ 
(roughly $10^{10}$ yr) starting from two
initial DM profiles (see Table I and text).
Colored curves: benchmark models; dashed lines: $\sigma v=0$;
dotted curves: $\sigma v=3\times 10^{-26}{\rm cm}^3{\rm s}^{-1}$,
$m=50$ GeV; thick line: initial DM density.
}
\label{fig1}
\end{figure}

We also carried out integrations for the set of benchmark models derived
in \cite{Battaglia:2003ab} in the framework of minimal 
supergravity (mSUGRA). Although other scenarios (supersymmetric or not) 
predict different parameters for the DM candidate, the values 
of $\sigma v/m$ are often approximately the same.
Light DM candidates \cite{Boehm:2003bt}, for example, 
have masses smaller than 20 MeV if they are to be responsible for the 511 keV emission
from the Galactic bulge \cite{Beacom:2004pe}, but they also 
typically have cross sections much smaller than the thermal cross 
section in the early universe, which implies that $\sigma v/m$ 
falls again in the same range discussed above. Heavy candidates,
like those proposed to explain the HESS data
\cite{Bergstrom:2004cy, Hooper:2004fh}, have masses in
the 10--20 TeV range, and thermal cross sections, so that they
fall again in the same range of $\sigma v/m$. 
Fig. 1 shows the final DM density profile for each of the
benchmark models, starting from DM models $N_{sp}$ and $A_{sp}$;
the latter model is the ``adiabatically contracted'' version 
of the former.
While adopting the maximal annihilation rate effectively
destroys the spike and produces $\overline{J}$ values as low as those
of spike-free models, 
other benchmark models with smaller $\sigma v/m$
result in strong enhancements in $\overline{J}$.

\begin{figure}[h]
\includegraphics[width=3.2in]{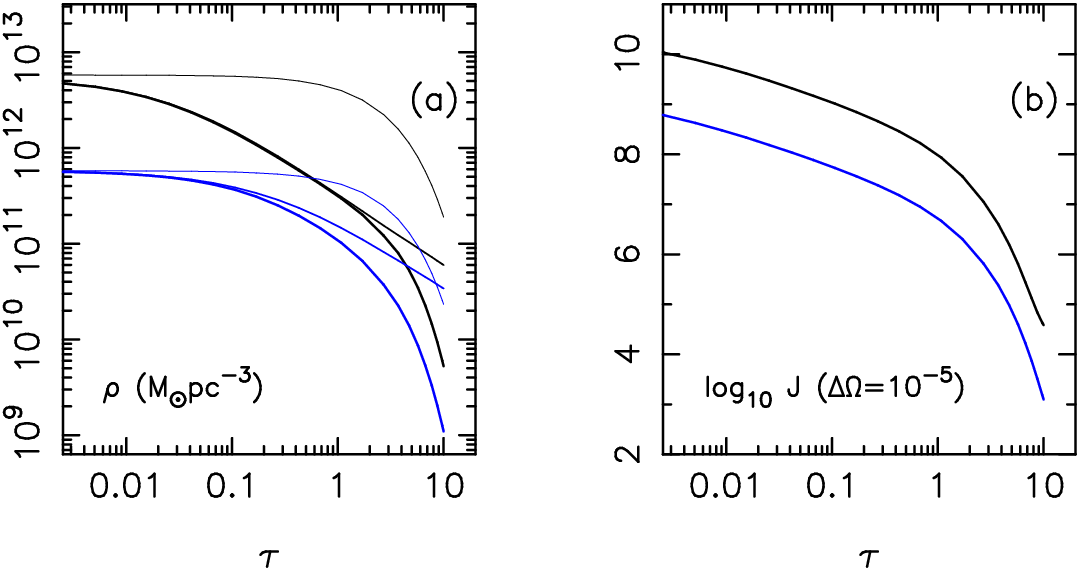}
\caption{(a) Evolution of the dark matter density at a radius
of $10^{-5}r_h\approx 2\times 10^{-5}{\rm pc}$ in a 
$\rho\sim r^{-2.33}$ spike, for $m=200$ GeV, 
and $\sigma v=10^{-27}{\rm cm}^3{\rm s}^{-1}$. 
The upper(lower) set of curves correspond to an initial 
density normalization
at $r_h$ of $10(100)M_\odot$ pc$^{-3}$.
In order of increasing thickness, the curves show the
evolution of $\rho$ in response to heating by stars;
to self-annihilations; and to both processes acting
together.
Time is in units of $T_{\rm heat}$ defined in the text;
$\tau=10$ corresponds roughly to $10^{10}$ yr.
(b) Evolution of $\overline{J}$ averaged over an angular
window of $10^{-5}$ sr.
}
\label{rhooft}
\end{figure}

Fig.~\ref{rhooft} shows the evolution of the dark matter density
at a radius of $10^{-5}r_h\approx 2\times 10^{-5}{\rm pc}$, 
starting from a $\rho\sim r^{-2.33}$ spike ($\rho\sim r^{-1}$ cusp).
Two values were taken for the initial density normalization
at $r=r_h$, $\rho(r_h)=(10,100)M_\odot{\rm pc}^{-3}$.
The self-annihilation term in Eq. (\ref{eq:fpa}) was computed
assuming $m=200$ GeV, 
$\sigma v=10^{-27}{\rm cm}^3{\rm s}^{-1}$.
The early evolution is dominated by self-annihilations
but for $t\gtrsim 10^9{\rm yr}\approx T_{\rm heat}$,
heating of dark matter by stars dominates.
The change in $\overline{J}_5$ (Fig. 2b.) is dramatic,
with final values in the range $10^3\lesssim\overline{J}\lesssim 10^5$.

We define the boost factor $b$ as $\overline{J}/\overline{J}_N$, 
with $\overline{J}$ the value in the evolved model
and $\overline{J}_N$ the value in a $\rho\propto r^{-1}$ 
(spike-free) halo with
the same density normalization at $r=R_\odot$.
Figure 2 shows boost factors at $\tau=10$ for each of the models
in Table I.
We found that the dependence of $B\equiv\log_{10}b$ on 
$\sigma v/m$
could be very well approximated by the function 
\begin{equation}
B(X)=B_{max}-(1/2)(B_{max}-B_{min})\{1+\tanh[C_1(X-C_2)]\}
\end{equation}
with $X\equiv\log_{10}(\sigma v/10^{-30}{\rm cm}^3{\rm s}^{-1})/
(m/100{\rm GeV})$.
Table II gives values of the fitting parameters at $\tau=10$ 
in each of the models with a spike.
The boost factor is independent of $\sigma v$ 
for low $\sigma v$, since annihilations
are unimportant in this limit, and also for high $\sigma v$, since
annihilations effectively destroy the spike.

\begin{table}
\caption{\label{Table2}
Parameters in the fitting function for the boost.
}
\begin{ruledtabular}
\begin{tabular}{ccccccccc}
 $\Delta\Omega=$ & & $10^{-3}$ & & & & $10^{-5}$ & \\
 & $B_{min}$ & $B_{max}$ & $C_1$ & $C_2$ & $B_{min}$ & $B_{max}$ & $C_1$ & $C_2$\\
$N_{sp}$ & $-0.02$ & $1.31$ & $0.66$ & $0.73$ & $-0.05$ & $2.35$ & $0.55$ & $1.50$ \\
$N_{c,sp}$ & $-0.02$ & $0.05$ & $0.75$ & $0.92$ & $-0.18$ & $0.38$ & $0.72$ & $1.31$ \\
$A_{sp}$ & $2.16$ & $6.29$ & $0.43$ & $-0.28$ & $2.97$ & $7.36$ & $0.41$ & $0.13$ \\
$A_{c,sp}$ & $1.96$ & $2.22$ & $0.74$ & $-0.49$ & $2.31$ & $3.00$ & $0.72$ & $-0.15$ \\
\end{tabular}
\end{ruledtabular}
\end{table}

\begin{figure}
\includegraphics[height=8cm]{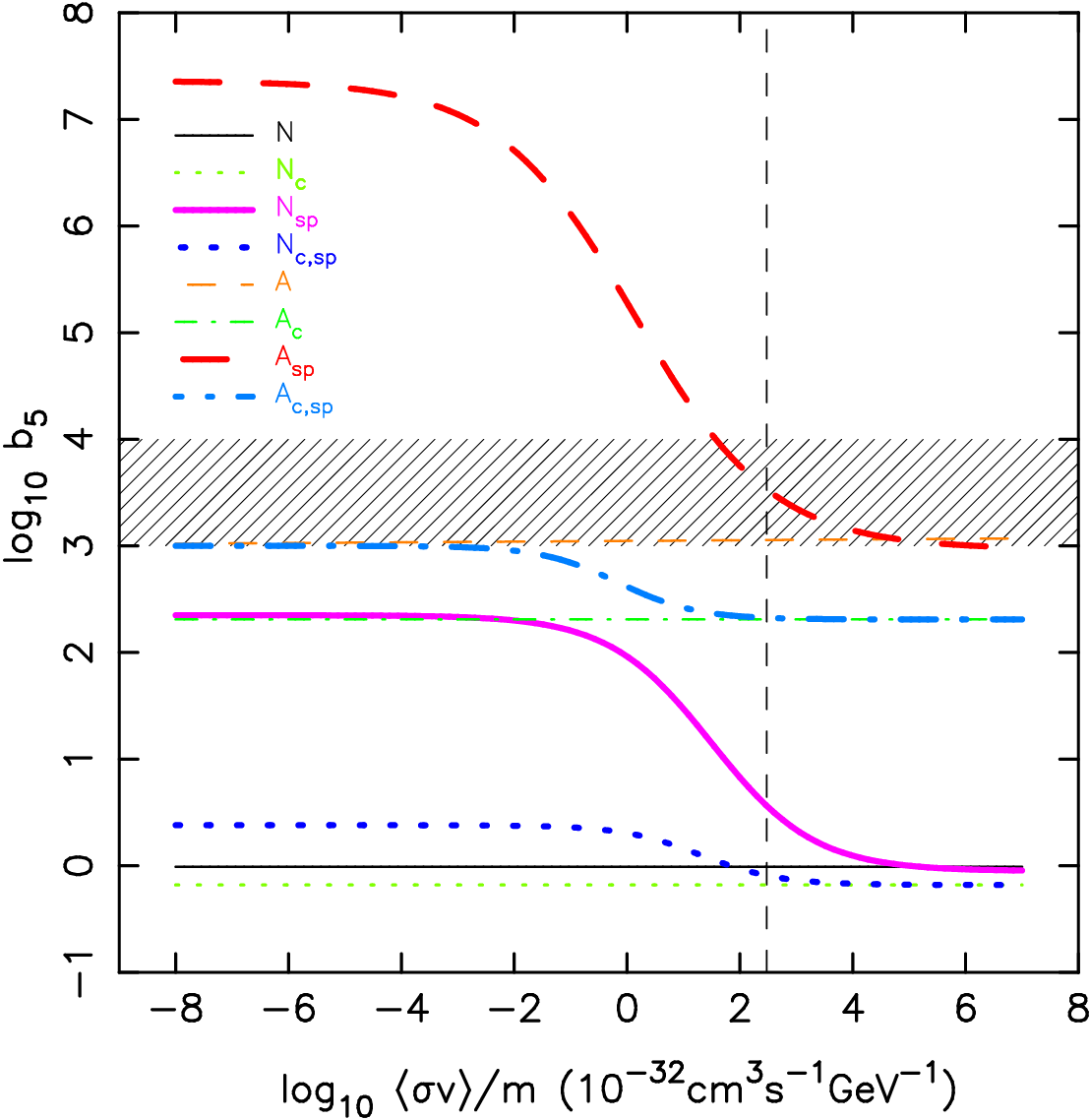}
\caption{Boost factors $b_5$ ($\Delta\Omega=10^{-5}$) as a function
of $\sigma v/m$ at $\tau=10$ for the DM models
of Table I.
Hatched region is the approximate boost factor required
to explain the HESS $\gamma$-ray detection if the 
particle mass is $\sim 10$ TeV and 
$\sigma v=3\times 10^{-26} \rm{cm}^3 \rm{s}^{-1}$
(vertical line).}
\label{fig2}
\end{figure}

We now apply these results to the study of high-energy $\gamma$-rays 
from dark matter annihilations at the GC.
An early detection by the EGRET collaboration, 
of a $\gamma$-ray source coincident with the position of the 
SBH~\cite{mayer}, has not been confirmed by a subsequent 
analysis~\cite{Hooper:2002ru}. 
However, Air Cherenkov Telescopes like HESS, CANGAROO and 
VERITAS have all detected a source coincident within their 
angular resolution with the GC SBH.
In particular the HESS data suggest a spectrum extending 
up to 10 TeV, with no apparent cut-off~\cite{Aharonian:2004wa}. 
It is difficult to interpret the observed emission as due to DM 
annihilation, since usual DM candidates are lighter 
than the required 10 TeV, and since the spectrum is quite
flat. The latter problem can be solved by considering 
processes like $\chi \chi \rightarrow \ell \bar{\ell} \gamma$
\cite{Bergstrom:2004cy}, where $\ell$ is a charged lepton, 
a channel heavily suppressed for
neutralinos, but open for Kaluza--Klein particles. Although
the contribution of the total flux is small (the channel is 
suppressed by a factor $\alpha / \pi$ with respect to the 
annihilation to charged leptons), the corresponding photon 
spectrum is very flat, with a sharp cut-off at an energy 
corresponding to the particle mass. The other problem, i.e.
the high dark matter particle mass required to reproduce the 
HESS data, can be solved in the framework of some specific
theoretical scenarios, such as those proposed 
in Refs.~\cite{Bergstrom:2004cy, Hooper:2004fh}).  
In this case, a boost factor of order $10^3\lap b\lap 10^4$ 
is required to match the observed normalization, 
for particle masses of order 10 TeV and cross sections of order 
$3\times 10^{-26} \rm{cm}^3 \rm{s}^{-1}$. 
Figure 2 shows that such boost factors are achievable in the 
adiabatically compressed DM models, $\rho\sim r^{-1.5}$,
especially if a spike is initially present, although the spike is
not required. 
We note that the particle models discussed above could 
easily evade the synchrotron constraints discussed in 
\cite{Bertone:2001jv,Gondolo:2000pn, Aloisio:2004hy}. 
Looking for example at Fig.~6 of Ref. \cite{Aloisio:2004hy}, 
we note that the synchrotron constraint is weaker for heavier masses, 
and the annihilation rate, in the case of the evolved $A_{sp}$ profile discussed above, is suppressed
by many orders of magnitude with respect to the case discussed 
in \cite{Aloisio:2004hy}, corresponding to a non-evolved $N_{sp}$ profile
(Table I).
A detailed analyis of indirect detection
of supersymmetric and Kaluza-Klein DM in light of 
this work will be presented elsewhere. 

This work was supported by grants AST-0071099, AST-0206031, AST-0420920 
and AST-0437519 from the NSF, grant NNG04GJ48G and NAG 5-10842 from NASA, 
and grant HST-AR-09519.01-A from STScI.

\end{document}